\def\mpl{m_{\cal P}}
\documentstyle[twoside,12pt]{article}

\catcode`\@=11
\long\def\@makefntext#1{ 
\protect\noindent \hbox to 3.2pt {\hskip-.9pt
{{\twelverm\@thefnmark}}$\hfil}#1\hfill} 

\def\thefootnote{\fnsymbol{footnote}}
\def\@makefnmark{\hbox to 0pt{$^{\@thefnmark}$\hss}}  

\def\ps@myheadings{\let\@mkboth\@gobbletwo
\def\@oddhead{\hbox{} 
\rightmark\hfil\twelverm\thepage}
\def\@oddfoot{}\def\@evenhead{\twelverm\thepage\hfil 
\leftmark\hbox{}}\def\@evenfoot{}
\def\sectionmark##1{}\def\subsectionmark##1{}}

\oddsidemargin=\evensidemargin
\newcommand{\symbolfootnote}{\renewcommand{\thefootnote}
	{\fnsymbol{footnote}}}
\renewcommand{\thefootnote}{\fnsymbol{footnote}}
\newcommand{\alphfootnote}
	{\setcounter{footnote}{0}
	 \renewcommand{\thefootnote}{\sevenrm\alph{footnote}}}

\newcounter{sectionc}\newcounter{subsectionc}\newcounter{subsubsectionc}
\renewcommand{\section}[1] {\vspace{14pt}\addtocounter{sectionc}{1}
\setcounter{subsectionc}{0}\setcounter{subsubsectionc}{0}\noindent
	{\twelvebf\thesectionc. #1}\par\vspace{7pt}}
\renewcommand{\subsection}[1] {\vspace{14pt}\addtocounter{subsectionc}{1}
	\setcounter{subsubsectionc}{0}\noinndent
	{\bf\thesectionc.\thesubsectionc. {\kern1pt \bfit #1}}\par\vspace{7pt}}
\renewcommand{\subsubsection}[1]
{\vspace{14pt}\addtocounter{subsubsectionc}{1}
	\noindent{\twelverm\thesectionc.\thesubsectionc.\thesubsubsectionc.
	{\kern1pt \twelveit #1}}\par\vspace{7pt}}
\newcommand{\nonumsection}[1] {\vspace{14pt}\noindent{\twelvebf #1}
	\par\vspace{7pt}}

\newcounter{appendixc}
\newcounter{subappendixc}[appendixc]
\newcounter{subsubappendixc}[subappendixc]
\renewcommand{\thesubappendixc}{\Alph{appendixc}.\arabic{subappendixc}}
\renewcommand{\thesubsubappendixc}
	{\Alph{appendixc}.\arabic{subappendixc}.\arabic{subsubappendixc}}

\renewcommand{\appendix}[1] {\vspace{12pt}
        \refstepcounter{appendixc}
        \setcounter{figure}{0}
        \setcounter{table}{0}
        \setcounter{lemma}{0}
        \setcounter{theorem}{0}
        \setcounter{corollary}{0}
        \setcounter{definition}{0}
        \setcounter{equation}{0}
        \renewcommand{\thefigure}{\Alph{appendixc}.\arabic{figure}}
        \renewcommand{\thetable}{\Alph{appendixc}.\arabic{table}}
        \renewcommand{\theappendixc}{\Alph{appendixc}}
        \renewcommand{\thelemma}{\Alph{appendixc}.\arabic{lemma}}
        \renewcommand{\thetheorem}{\Alph{appendixc}.\arabic{theorem}}
        \renewcommand{\thedefinition}{\Alph{appendixc}.\arabic{definition}}
        \renewcommand{\thecorollary}{\Alph{appendixc}.\arabic{corollary}}
        \renewcommand{\theequation}{\Alph{appendixc}.\arabic{equation}}
        \noindent{\twelvebf Appendix \theappendixc #1}\par\vspace{5pt}}
\newcommand{\subappendix}[1] {\vspace{12pt}
        \refstepcounter{subappendixc}
        \noindent{\bf Appendix \thesubappendixc. {\kern1pt \bfit #1}}
	\par\vspace{5pt}}
\newcommand{\subsubappendix}[1] {\vspace{12pt}
        \refstepcounter{subsubappendixc}
        \noindent{\rm Appendix \thesubsubappendixc. {\kern1pt \twelveit #1}}
	\par\vspace{5pt}}

\newcommand{\textlineskip}{\baselineskip=14pt}
\newcommand{\smalllineskip}{\baselineskip=12pt}

\def\eightcirc{
\begin{picture}(0,0)
\put(4.4,1.8){\circle{6.5}}
\end{picture}}
\def\eightcopyright{\eightcirc\kern2.7pt\hbox{\eightrm c}}

\newcommand{\copyrightheading}[1]
	{\vspace*{-2.5cm}\smalllineskip{\flushleft
	{\twelverm International Journal of Modern Physics D, #1}\\
	{\twelverm $\eightcopyright$\, World Scientific Publishing
	 Company}\\
	 }}

\def\abstracts#1#2#3{{
	\centering{\begin{minipage}{5.5in}\baselineskip=14pt\twelverm
	\centerline{ABSTRACT}
        \vspace{0.2truein}
	\parindent=0pt #1\par
	\parindent=15pt #2\par
	\parindent=15pt #3
	\end{minipage} }\par}}

\def\keywords#1{{
	\centering{\begin{minipage}{4.5in}\baselineskip=14pt\tenrm
	{\twelveit Keywords}\/: #1
	 \end{minipage} }\par }}

\newcommand{\bibit}{\nineit}
\newcommand{\bibbf}{\ninebf}
\renewenvironment{thebibliography}[1]			
	{\twelverm
 \baselineskip=14pt				
	 \begin{list}{\arabic{enumi}.}
	{\usecounter{enumi}\setlength{\parsep}{0pt}
	 \setlength{\leftmargin 17pt}{\rightmargin 0pt}	
	 \setlength{\itemsep}{0pt} \settowidth		
	{\labelwidth}{#1.}\sloppy}}{\end{list}}

\newcounter{itemlistc}
\newcounter{romanlistc}
\newcounter{alphlistc}
\newcounter{arabiclistc}

\newcommand{\fcaption}[1]{
        \refstepcounter{figure}
        \setbox\@tempboxa = \hbox{\tenrm Fig.~\thefigure. #1}
        \ifdim \wd\@tempboxa > 5in
           {\begin{center}
        \parbox{5in}{\tenrm \smalllineskip Fig.~\thefigure. #1 }
            \end{center}}
        \else
             {\begin{center}
             {\tenrm Fig.~\thefigure. #1}
              \end{center}}
        \fi}

\newcommand{\tcaption}[1]{
        \refstepcounter{table}
        \setbox\@tempboxa = \hbox{\tenrm Table~\thetable. #1}
        \ifdim \wd\@tempboxa > 5in
           {\begin{center}
        \parbox{5in}{\tenrm\smalllineskip Table~\thetable. #1 }
            \end{center}}
        \else
             {\begin{center}
             {\tenrm Table~\thetable. #1}
              \end{center}}
        \fi}

\def\@citex[#1]#2{\if@filesw\immediate\write\@auxout	
	{\string\citation{#2}}\fi			
\def\@citea{}\@cite{\@for\@citeb:=#2\do			
	{\@citea\def\@citea{,}\@ifundefined		
	{b@\@citeb}{{\bf ?}\@warning
	{Citation `\@citeb' on page \thepage \space undefined}}
	{\csname b@\@citeb\endcsname}}}{#1}}

\newif\if@cghi
\def\cite{\@cghitrue\@ifnextchar [{\@tempswatrue
	\@citex}{\@tempswafalse\@citex[]}}
\def\citelow{\@cghifalse\@ifnextchar [{\@tempswatrue
	\@citex}{\@tempswafalse\@citex[]}}
\def\@cite#1#2{{$\null^{#1}$\if@tempswa\typeout
	{IJCGA warning: optional citation argument
	ignored: `#2'} \fi}}
\newcommand{\citeup}{\cite}

\def\pmb#1{\setbox0=\hbox{#1}
	\kern-.025em\copy0\kern-\wd0
	\kern.05em\copy0\kern-\wd0
	\kern-.025em\raise.0433em\box0}
\def\mbi#1{{\pmb{\mbox{\scriptsize ${#1}$}}}}
\def\mbr#1{{\pmb{\mbox{\scriptsize{#1}}}}}

\def\fnm#1{$^{\mbox{\scriptsize #1}}$}
\def\fnt#1#2{\footnotetext{\kern-.3em
	{$^{\mbox{\scriptsize #1}}$}{#2}}}

\def\fpage#1{\begingroup
\voffset=.3in
\thispagestyle{empty}\begin{table}[b]\centerline{\footnotesize #1}
	\end{table}\endgroup}

\def\runninghead#1#2{\pagestyle{myheadings}
\markboth{{\tenit{\quad #1}}\hfill}{\hfill{\tenit{#2\quad}}}}

\font\it=cmti12
\font\bf=cmbx12
\font\twelvebf=cmbx12
\font\twelveit=cmti12
\font\twelveit=cmti12
\font\bfit=cmbxti10 at 10pt
\font\twelvebf=cmbx12 
\font\twelverm=cmr12  
\font\twelveit=cmti12 
\font\tenbf=cmbx10
\font\tenrm=cmr10
\font\tenit=cmti10
\font\tenit=cmti10
\font\ninerm=cmr9
\font\sevenrm=cmr7

\newcommand{\proof}[1]{{\tenbf Proof.} #1 $\Box$.}

\def\qed{\hbox{${\vcenter{\vbox{                          
   \hrule height 0.4pt\hbox{\vrule width 0.4pt height 6pt
   \kern5pt\vrule width 0.4pt}\hrule height 0.4pt}}}$}}

\begin{document}
\normalsize\textlineskip
{\thispagestyle{empty}
\setcounter{page}{1}
\pagestyle{empty}

\renewcommand{\thefootnote}{\fnsymbol{footnote}} 

%
%

\centerline{\hfill ALBERTA THY/07-95}
\vspace{0.7truein}
\centerline{\bf THE NATURE OF COSMIC TIME }
\vspace{1.0truein}
\centerline{\bf D.S. SALOPEK}
\vspace{0.2truein}
\centerline{\it Department of Physics}
\centerline{\it University of Alberta}
\centerline{\it Edmonton, Canada T6G 2J1 }
%
\vspace*{1.0truein}
\abstracts{
\noindent
Hamilton-Jacobi theory provides a natural
starting point for a covariant description of the gravitational field.
Using a spatial gradient expansion, one may solve
for the phase of the wavefunction by using a line-integral
in superspace. Each contour of integration corresponds to a particular
choice of time-hypersurface, and each yields
the same answer. In this way, one can describe all time
choices simultaneously. In an interesting application to
cosmology, I compute large-angle microwave
background anisotropies and the galaxy-galaxy correlation
function associated with the scalar and tensor fluctuations
of power-law inflation. }{}{}
\vspace{1.0truein}
\centerline{Fields Institute Publication}
\vspace{0.1truein}
\centerline{\it Sixth Canadian Conference on}
\centerline{\it General Relativity and Relativistic Astrophysics}
\vspace{0.1truein}
\centerline{May 25-27, 1995}
\vfill\eject
\vspace*{-3pt}\textlineskip
\section{Introduction}
\noindent
Hamilton-Jacobi (HJ) theory is a cornerstone of modern
theoretical physics. It may be profitably applied
to numerous problems in cosmology. Since a full quantum theory is lacking,
a semi-classical analysis provides our best understanding
of the gravitational field.

HJ theory provides an elegant formalism for computing
density perturbations as well as microwave background
fluctuations arising from the inflationary
scenario.\cite{SS95}{}$^{,}$\cite{HH85}
It has also been successfully employed in deriving the Zel'dovich
approximation\cite{Zel} (which describes the formation
of sheet-like structures in the Universe)
from general relativity.\cite{CPSS94}
Numerous researchers have employed HJ methods in an attempt
to recover the inflaton potential
from cosmological observations.\cite{COPELAND93} Lastly,
HJ techniques can be used to construct inflationary models that
yield non-Gaussian primordial fluctuations;\cite{SB1} such models
could possibly resolve the problems of large scale structure
in the Universe.\cite{Mosc93}

I will focus on one particularly attractive feature of
HJ theory: it provides a covariant formulation of the
gravitational field.\cite{PSS94} In the semi-classical theory, the answer
to the question of time is clear: time is arbitrary.
HJ theory enables one to consider all such time choices
simultaneously. I will now consider a simple analogy from potential theory
which illuminates the general technique.

\section{Potential Theory}

The fundamental problem in potential theory is: given a force
field $g^i(u_k)$ which is a function of $n$ variables $u_k$,
what is the potential $\Phi \equiv \Phi(u_k)$ (if it exists)
whose gradient returns the force field,
\begin{equation}
{\partial \Phi \over \partial u_i} = g^i(u_k) \quad ?
\end{equation}
Not all force fields are derivable from
a potential. Provided that the force field satisfies the
integrability relation,
\begin{equation}
0= {\partial g^i \over \partial u_j} - {\partial g^j \over \partial u_i} =
\left [{\partial  \over \partial u_j}, {\partial  \over  \partial u_i }
\right ] \, \Phi \, ,
\end{equation}
(i.e., it is curl-free),
one may find a solution which is conveniently expressed using a
line-integral
\begin{equation}
\Phi(u_k) = \int_C \sum_j dv_j \ g^j(v_l) \ .
\end{equation}
If the two endpoints are fixed, all contours return the same
answer. In practice, one employs the simplest contour that
one can imagine: a line connecting the origin to the
observation point $u_k$. Using $s$, $0 \le s \le 1$,
to parameterize the contour, the line-integral may be rewritten as
\begin{equation}
\Phi(u_k) = \sum_{j=1}^n \int_0^1  ds  \; u_j \ g^j(su_k) \ .
\label{lint}
\end{equation}
Similarly, in solving for the phase of the wavefunctional,
one utilizes a line-integral in {\it superspace}.

\section{Solving the Hamilton-Jacobi Equation for General Relativity}
\noindent
The Hamilton-Jacobi equation for general relativity is
derived using a Hamiltonian formulation of gravity.
One first writes the line element using the ADM 3+1 split,
\begin{equation}
ds^2=\left(-N^2+\gamma^{ij}N_iN_j\right)dt^2 + 2N_idt\,dx^i + \gamma_{ij} \,\
dx^i dx^j\ ,
\label{ADMdecomp}
\end{equation}
where $N$ and $N_i$ are the lapse and shift functions, respectively,
and $\gamma_{ij}$ is the 3-metric. Hilbert's action for gravity interacting
with a scalar field becomes
\begin{equation}
{\cal I}=\int d^4x\left(\pi^{\phi}\dot\phi +\pi^{ij}\dot\gamma_{ij}
-N{\cal H} -N^i{\cal H}_i\right).
\label{ADMaction}
\end{equation}
The lapse and shift functions are Lagrange multipliers that
imply the energy constraint ${\cal H}(x)=0$ and the
momentum constraint ${\cal H}_i(x)=0$.

The object of chief importance is the generating functional
${\cal S}\equiv {\cal S}[\gamma_{ij}(x), \phi(x)]$.
For each universe with field configuration
$[\gamma_{ij}(x), \phi(x)]$ it assigns a number
which can be complex. The generating functional is
the `phase' of the wavefunctional in the semi-classical approximation:
$\Psi \sim e^{i{\cal S}}$. The probability functional,
${\cal P} \equiv |\Psi|^2$, is given by the square of the wavefunctional.

Replacing the conjugate momenta by functional derivatives
of ${\cal S}$ with respect to the fields,
\begin{equation}
\pi^{ij}(x)={\delta{\cal S}\over \delta{\gamma_{ij}(x)}}\ , \qquad
\pi^{\phi}(x)={\delta{\cal S}\over \delta\phi (x)}\ ,
\label{pis}
\end{equation}
and substituting into the energy constraint,
one obtains the Hamilton-Jacobi equation,
\begin{eqnarray}
{\cal H}(x)=&&\gamma^{-1/2} {\delta{\cal S}\over \delta\gamma_{ij}(x)}
{\delta{\cal S}\over \delta\gamma_{kl}(x)}
\left[2\gamma_{il}(x) \gamma_{jk}(x) - \gamma_{ij}(x)\gamma_{kl}(x)\right]
\nonumber \\
&& + {1\over 2} \gamma^{-1/2}
\left({\delta{\cal S}\over \delta\phi(x)}\right)^2
+\gamma^{1/2}V(\phi(x)) \nonumber \\
&& -{1\over 2}\gamma^{1/2}R
+{1\over 2} \gamma^{1/2}\gamma^{ij}\phi_{,i}\phi_{,j}=0 \ ,
\label{HJequation}
\end{eqnarray}
which describes how ${\cal S}$ evolves in superspace.
$R$ is the Ricci scalar associated with the 3-metric, and $V(\phi)$
is the scalar field potential.
In addition, one must also satisfy the momentum constraint
\begin{equation}
{\cal H}_{i}(x)=-2\left(\gamma_{ik}{\delta{\cal S}\over \delta\gamma_{kj}(x)}
\right)_{,j} +
{\delta{\cal S}\over\delta\gamma_{lk}(x)}\gamma_{lk,i} +
{\delta{\cal S}\over\delta\phi (x)} \phi_{,i}=0 \ ,
\label{Smomentum}
\end{equation}
which legislates gauge invariance: ${\cal S}$ is invariant under
reparametrizations of the spatial coordinates.\cite{Peres}
(Units are chosen so that
$c=8\pi G= \hbar= 1$). Since neither the lapse function
nor the shift function appears in
eqs.(\ref{HJequation},\ref{Smomentum}) the temporal
and spatial coordinates are {\it arbitrary}:
HJ theory is {\it covariant}.

As a first step in solving eqs.(\ref{HJequation},\ref{Smomentum}),
I will expand the generating functional
\begin{equation}
{\cal S}= {\cal S}^{(0)} + {\cal S}^{(2)} + {\cal S}^{(4)} + \dots\ ,
\label{theexpansion}
\end{equation}
in a series of terms according to
the number of spatial gradients that they contain.
The invariance of the generating functional
under spatial coordinate transformations
suggests a solution of the form,
\begin{equation}
{\cal S}^{(0)}[ \gamma_{ij}(x), \phi(x)] = - 2 \int d^3x
\gamma^{1/2} H \left[ \phi(x) \right] \ ,
\end{equation}
for the zeroth order term ${\cal S}^{ (0) }$. The function
$H \equiv H(\phi)$ satisfies the separated HJ equation of
order zero,\cite{SB1}
\begin{equation}
H^2={2\over 3}\left( {\partial H\over\partial\phi} \right)^2
 +{1 \over 3}V\left( \phi \right ) \ ,
\label{Hequation}
\end{equation}
which is an ordinary differential equation.
Note that ${\cal S}^{(0)}$ contains no spatial gradients.

In order to compute the higher order terms, one introduces a
change of variables, $( \gamma_{ij}, \phi ) \rightarrow
(f_{ij}, u)$:
\begin{equation}
u = \int \  { d \phi \over -2 { \partial H \over \partial \phi } } \ ,
\quad f_{ij} = \Omega^{-2}(u) \, \gamma_{ij} \ , \label{changeA}
\end{equation}
where the conformal factor $\Omega \equiv \Omega(u)$ is defined through
\begin{equation}
{ d \ln \Omega \over d u} \equiv -2 { \partial H \over \partial \phi}
{ \partial \ln \Omega \over \partial \phi} = H \ . \label{changeB}
\end{equation}
in which case the equation for ${\cal S}^{(2m)}$ becomes
\begin{equation}
{\delta {\cal S}^{(2m)}\over\delta u(x)}\Bigg|_{f_{ij}}
+ {\cal R}^{(2m) }[u(x), f_{ij}(x)] =0\ .
\label{HJ.conf.for.dust}
\end{equation}
The remainder term ${\cal R}^{(2m) }$ depends on some quadratic
combination of the previous order terms ({\it i.e.}, it may be written
explicitly\cite{PSS94}). For example, for $m=1$, it is
\begin{equation}
{\cal R}^{(2)} =
{1\over 2} \gamma^{1/2}\gamma^{ij}\phi_{,i}\phi_{,j}
-{1\over 2}\gamma^{1/2}R \, .
\end{equation}
Eq.(\ref{HJ.conf.for.dust}) has the form of an infinite dimensional gradient.
It may integrated using a line integral analogous to
eq.(\ref{lint}):
\begin{equation}
{\cal S}^{(2m)}=-\int d^3x\int_0^1  ds\  u(x) \
{\cal R}^{(2m)}[su(x), f_{ij}(x)] \; .
\label{lints}
\end{equation}
Typically, ${\cal S}^{(2m)}$ is an integral of terms which contain
the Ricci tensor and derivatives of the scalar field.\cite{PSS94}

The integrability
condition for the HJ equation\cite{MT72} follows from the
Poisson bracket of the energy constraints evaluated at
spatial points $x$ and $x^\prime$,
\begin{equation}
\{ {\cal H}(x^k), {\cal H}(x^{k^\prime}) \} =
[ \gamma^{ij}(x^k) {\cal H}_j(x^k)+
\gamma^{ij}(x^{k^\prime}) {\cal H}_j(x^{k^\prime}) ]
\; { \partial \over \partial x^i} \delta^3(x^k - x^{k^\prime}) \, .
\label{poisson}
\end{equation}
In fact, alternative contours replacing the
line-integral eq.(\ref{lints}) will
correspond to different time-hypersurface choices.
Provided that the generating functional is invariant
under reparametrizations of the spatial coordinates,
(e.g., ${\cal H}_i$ vanishes in the right-hand-side of
eq.(\ref{poisson})),
different time-hypersurface choices will lead to the same
generating functional. Hypersurface invariance is closely
related to gauge invariance.

\section{Computing Large-Angle Microwave Background Fluctuations
and Galaxy Correlations}
\noindent
In order to describe the fluctuations arising during the
inflationary epoch, it is necessary to sum an infinite
subset\cite{SS95} of the terms ${\cal S}^{(2m)}$.
In this case, one considers all terms which are quadratic in the Ricci
tensor $\tilde R_{ij}$ of the conformal 3-metric $f_{ij}(x)$
defined in eq.(\ref{changeA}).
Once again, no explicit choice of time hypersurface is made.

However, when one compares theory with observations, there
are indeed preferred gauges. The phase transition
of photon-decoupling occurs essentially on a uniform
temperature slice, $T \sim 4000 K$, when protons
combine with electrons to form neutral hydrogen.  For adiabatic
perturbations at large wavelengths, this slice
is the same as a comoving, synchronous time hypersurface
which Sachs and Wolfe\cite{SW67} used in the computation
of large-angle microwave background anisotropies.

The power-law inflationary model\cite{LM85}
provides an excellent example of HJ techniques.
For this model, the scalar factor of the Universe evolves as
$a \sim t^p$ which describes an inflationary epoch provided $p> 1$.
The scalar field potential has an exponential form
\begin{equation}
V(\phi)= V_0 \; {\rm exp} \left(-\sqrt{2\over p} \phi \right) \; .
\label{s.f.potential}
\end{equation}
Power-law inflation is of high interest for observational
cosmology because it may produce
copious amounts of primordial gravitational
radiation,\cite{S92}{}$^,$\cite{S95}
which is in essence a quantum gravitational effect.

\begin{figure}[htbp]
\setlength{\unitlength}{0.240900pt}
\ifx\plotpoint\undefined\newsavebox{\plotpoint}\fi
\sbox{\plotpoint}{\rule[-0.175pt]{0.350pt}{0.350pt}}%

\fcaption{
For the present epoch, the power spectra
for the linear density perturbation $\delta \rho/ \rho$ is
shown. The data points are the observed power spectrum derived
from galaxy surveys. The curves are theoretical predictions
of the power-law inflationary model for several values of $p$:
$p=\infty$ is the standard cold-dark-matter model;
$p=21$ provides the best fit.}
\end{figure}

Fig.(1) illustrates the observed power spectrum,
\begin{equation}
{\cal P}_\delta(k) \equiv { k^3 \over 2 \pi^2} \int d^3x \;
e^{-i \vec k \cdot \vec x} \;
<{\delta \rho(x) \over \rho } {\delta \rho(0) \over \rho } > \, ,
\end{equation}
for the linear density perturbation at the present
epoch; here $k$ is the comoving wavenumber. The data points were
compiled using eight galaxy surveys.\cite{PD94}
Also shown are the power spectra arising from
power-law inflation for various values of $p$. After the inflationary epoch,
I have assumed that the evolution of the fluctuations is
described by the cold-dark-matter transfer function\cite{PEEBLES82}
where the present Hubble parameter is taken to be
$H_0= 50 \;{\rm km \; s}^{-1} {\rm Mpc}^{-1}$.
With a correction for gravitational waves, the theoretical power spectra
for density perturbations have been normalized using the
2-year DMR data set\cite{BENNETT94}
of the Cosmic Background Explorer (COBE) satellite:
$\sigma_{sky}(10^0)= 30.5 \pm 2.7 \mu K \quad (68\% \;
{\rm confidence \;level})$.

The discrepancy between the galaxy surveys and
the standard cold-dark-matter model
($p= \infty$) is quite severe at short length scales,
$k > 10^{-1.4} \; {\rm Mpc}^{-1}$.
The bold line, $p=21$, provides the best fit to the observed data.
The agreement is excellent at short scales. At longer scales,
the theoretical model under-predicts the observed power but
the deficit is not very severe. For $p=21$, gravitational waves
contribute 35$\%$ to $\sigma^2_{sky}$, the
square of COBE's microwave anisotropy.

\section{Summary}
\noindent
The question of time choice in general relativity is a
difficult one, particularly for the quantum theory.\cite{KUCHAR}
For semi-classical problems of interest to observational cosmology,
one may construct a covariant formalism which treats
all time choices on an equal footing.
Power-law inflation with $p=21$ yields a better fit
to cosmological data than the standard cold-dark-matter model.

\section{Acknowledgments}

\noindent
I thank J.M. Stewart, J. Parry and K.M. Croudace for a fruitful
collaboration on Hamilton-Jacobi topics. This work was supported
by NSERC and CITA of Canada.

\nonumsection{References}

\vfill\eject

\end{document}